\documentclass[aps,prl,twocolumn,showpacs,amsmath,amssymb]{revtex4-2}
\usepackage{amsmath}
\usepackage{graphicx}
\usepackage{subfigure}
\usepackage{epstopdf}
\usepackage{color}
\usepackage{multirow}
\usepackage{setspace}
\usepackage{overpic}
\usepackage{amssymb}
\usepackage[bookmarksnumbered, pdfstartview=FitH,colorlinks,urlcolor=blue, citecolor=blue,linkcolor=blue] {hyperref}
\usepackage{lineno}
\usepackage{bm}
\usepackage{rotating}
\usepackage[utf8]{inputenc}

 \newcommand{\BESIIIorcid}[1]{\href{https://orcid.org/#1}{\hspace*{0.1em}\raisebox{-0.45ex}{\includegraphics[width=1em]{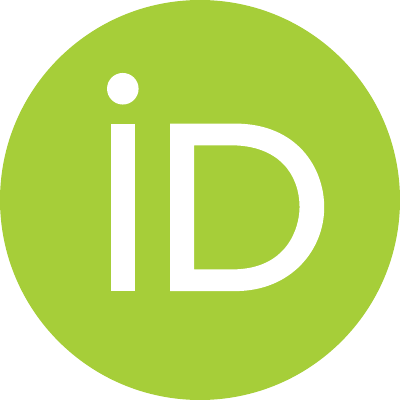}}}}

\let\oldequation\equation
\let\oldendequation\endequation

\renewenvironment{equation}
  {\linenomathNonumbers\oldequation}
  {\oldendequation\endlinenomath}

\begin{document}

\title{\bf \boldmath First Observation and Dynamical Study of the $D^+_s\to f_{0}(980) \mu^+\nu_\mu$ Decay}

\author{
M.~Ablikim$^{1}$\BESIIIorcid{0000-0002-3935-619X},
M.~N.~Achasov$^{4,c}$\BESIIIorcid{0000-0002-9400-8622},
P.~Adlarson$^{83}$\BESIIIorcid{0000-0001-6280-3851},
X.~C.~Ai$^{89}$\BESIIIorcid{0000-0003-3856-2415},
C.~S.~Akondi$^{31A,31B}$\BESIIIorcid{0000-0001-6303-5217},
R.~Aliberti$^{39}$\BESIIIorcid{0000-0003-3500-4012},
A.~Amoroso$^{82A,82C}$\BESIIIorcid{0000-0002-3095-8610},
Q.~An$^{78,65,\dagger}$,
Y.~H.~An$^{89}$\BESIIIorcid{0009-0008-3419-0849},
M.~S.~Anderson$^{39}$\BESIIIorcid{0009-0008-1550-2632}
% ...
\\
\vspace{0.2cm}
(BESIII Collaboration)\\
\vspace{0.2cm} {\it
$^{1}$ Institute of High Energy Physics, Beijing 100049, People's Republic of China\\
% ...
$^{4}$ Budker Institute of Nuclear Physics SB RAS (BINP), Novosibirsk 630090, Russia\\
% ...
$^{31}$ INFN Sezione di Ferrara, (A)INFN Sezione di Ferrara, I-44122, Ferrara, Italy; (B)University of Ferrara, I-44122, Ferrara, Italy\\
% ...
$^{39}$ Johannes Gutenberg University of Mainz, Johann-Joachim-Becher-Weg 45, D-55099 Mainz, Germany\\
% ...
$^{65}$ State Key Laboratory of Particle Detection and Electronics, Beijing 100049, Hefei 230026, People's Republic of China\\
% ...
$^{78}$ University of Science and Technology of China, Hefei 230026, People's Republic of China\\
% ...
$^{82}$ University of Turin and INFN, (A)University of Turin, I-10125, Turin, Italy; (B)University of Eastern Piedmont, I-15121, Alessandria, Italy; (C)INFN, I-10125, Turin, Italy\\
$^{83}$ Uppsala University, Box 516, SE-75120 Uppsala, Sweden\\
% ...
$^{89}$ Zhengzhou University, Zhengzhou 450001, People's Republic of China\\
\vspace{0.2cm}
$^{\dagger}$ Deceased\\
% ...
$^{c}$ Also at the Novosibirsk State University, Novosibirsk, 630090, Russia\\
% ...
}}


\begin{abstract}
Using 7.33 fb$^{-1}$ of $e^+e^-$ annihilation data recorded with the BESIII detector at center-of-mass energies from 4.128 to 4.226 GeV,
we report the first observation and dynamical study of the semileptonic decay $D^+_s\to f_{0}(980) \mu^+\nu_\mu$.
The absolute branching fraction of $D^+_s\to f_{0}(980) \mu^+\nu_\mu$ with $ f_{0}(980)\to \pi^+ \pi^-$ is $(1.59 \pm 0.18_{\rm stat} \pm 0.11_{\rm syst}) \times10^{-3}$.
Combining this result with our earlier BESIII measurement of ${\mathcal B}(D^+_s\to f_{0}(980) e^+\nu_e)$, their ratio is found to be $\frac{{\mathcal B}(D^+_s\to f_{0}(980) \mu^+\nu_\mu)}{{\mathcal B}(D^+_s\to f_{0}(980)e^+\nu_e)} = 0.92\pm0.13_{\rm stat}\pm0.08_{\rm syst}$, in agreement with the Standard Model expectation of lepton flavor universality.
From a dynamical analysis of the $D_{s}^{+} \to f_{0}(980)\mu^+\nu_\mu$ decay with a simple pole parametrization for the hadronic transition form factor, the product of the form factor $f^{f_{0}(980)}_{+}(0)$ and the $c\to s$ Cabibbo-Kobayashi-Maskawa matrix element $|V_{cs}|$ is determined to be $f^{f_{0}(980)}_{+}(0)|V_{cs}|=0.490\pm0.059_{\rm stat}\pm0.025_{\rm syst}$.
Averaging with our previously reported result for the $D_{s}^{+} \to f_{0}(980)e^+\nu_e$ decay, we obtain $f^{f_{0}(980)}_{+}(0)|V_{cs}|=0.500\pm0.016_{\rm stat}\pm0.020_{\rm syst}$. Using $|V_{cs}|$ from the CKMfitter group, we extract $f^{f_{0}(980)}_{+}(0)=0.514\pm0.017_{\rm stat}\pm0.021_{\rm syst}$. This represents the most precise determination of the $D_{s} \to f_{0}(980)$ transition form factor to date, and provides stringent tests of various theoretical models.

\end{abstract}

\maketitle

\oddsidemargin  -0.2cm
\evensidemargin -0.2cm

The nature of the light scalar mesons $f_0(980)$ and $a_0(980)$ has long remained controversial since their discovery more than half a century ago~\cite{pdg, Pelaez2016,weiwang2010,Jaffe1977}, owing to the non-perturbative nature of QCD at low energies. The existence of the radiative decay $\phi \to \gamma f_0(980)$~\cite{pdg} indicates a substantial $s\bar{s}$ component in this state. Nevertheless, the strong coupling of $f_0(980)$ to the $\pi\pi$ and $K\bar{K}$ thresholds complicates traditional spectroscopic analyses~\cite{FF-rate-sq2} and hinders a full description of its properties within a pure $ s\bar{s}$ scenario.
The mass ordering of these light scalar mesons cannot be reproduced by a conventional $q\bar{q}$ picture~\cite{pdg,qqbar1,qqbar2,QCDSR1,QCDSR2,LCSR1,LCSR2,LF-RQM,Soni2020}.  Therefore, exotic configurations, including diquark-antidiquark states~\cite{tetraquark1,tetraquark2,tetraquark3} and meson-meson molecular states~\cite{molecule1,molecule2,molecule3} have been proposed to reconcile its observed characteristics. Further experimental data are essential for clarifying the nature of the $f_0(980)$ scalar resonance.

Semileptonic (SL) decays of charmed mesons provide a unique and clean probe for investigating light scalar mesons. Since the final state leptons and hadrons are coupled only through the weak interaction, transition form factors (FFs) can be extracted, enabling the exploration of the non-perturbative Quantum Chromodynamics (QCD) mechanisms in the hadronic part of the decay. Moreover, given that the transition FFs and BFs are highly sensitive to the internal structure of light scalar states, the dynamical study of $D_s^{+} \to f_{0}(980) \ell^+ \nu_\ell$ probes the intrinsic nature of the $f_0(980)$~\cite{Achasov2012,Oset2015,Soni2020}. Experimentally, the CLEO Collaboration reported a BF measurement of \(D_{s}^{+} \to f_{0}(980)e^{+}\nu_{e}\) with \(f_{0}(980)\to \pi^{+}\pi^{-}\)~\cite{cleo2009-Ds2semilep,cleo2009-Ds2f0enu,cleo2015-Ds2semilep}. Recently, the BESIII Collaboration measured the BF of this process with higher precision and also extracted of the FF $f^{f_{0}(980)}_{+}(0)$~\cite{bes3f0980}; they also performed an independent BF measurement using \(D_{s}^{+} \to f_{0}(980)e^{+}\nu_{e}\) with \(f_{0}(980)\to \pi^{0}\pi^{0}\)~\cite{bes3-Ds2f0enu}. However, all previous measurements are restricted to the semi-electronic channel. The semi-muonic decay, $D_{s}^{+} \to f_{0}(980) \mu^+ \nu_\mu$, has not yet been observed. Its first measurement would provide an important complement and aid studies of the structure of the $f_{0}(980)$.

SL $D_{s}$ decays also offer an excellent opportunity to test lepton flavor universality~(LFU), which is a fundamental prediction of the Standard Model (SM). Refs.~\cite{prd91_094009,cpc_063107} point out that observable LFU violation effects may exist in SL decays induced by the $c\rightarrow s \ell \nu$ transition. All experimental tests of LFU in SL charmed meson decays have so far focused solely on final-state pseudoscalar and vector mesons, yielding results consistent with the SM.
In contrast, an analogous test of LFU in decays to scalar mesons dominated by the $c \to s$ transition awaits the measurement of the semi-muonic $D_s$ decay channels. 
In the SM, the ratio $\frac{\mathcal{B}(D_{s}^{+} \to f_{0}(980) \mu^+ \nu_\mu)}{\mathcal{B}(D_{s}^{+} \to f_{0}(980) e^+ \nu_e)}$ is predicted to be in the range of $0.94-1.11$~\cite{QCDSR2,LCSR1,LCSR2,LF-RQM,Soni2020}. There also exist significantly differing theoretical predictions for the BF of $D_{s}^{+} \to f_{0}(980) \mu^+ \nu_\mu$~\cite{LCSR2,LF-RQM,Soni2020}, with values spanning the range of ($0.14-0.26$)\%. Therefore, a measurement of $D_{s}^{+} \to f_{0}(980) \mu^+ \nu_\mu$ decay  is crucial for distinguishing among these theoretical models and extending tests of $\mu-e$ LFU with the SL decays.

In this Letter,  we report the first observation of the decay $D_{s}^{+} \to f_{0}(980) \mu^+ \nu_\mu$ with $f_{0}(980)\to \pi^{+}\pi^{-}$, providing an absolute BF measurement and a study of its dynamics. These measurements are performed using an $e^+e^-$ annihilation data sample corresponding to an integrated luminosity of 7.33 fb$^{-1}$ produced at center-of-mass energies ($E_{\rm CM}$)  between 4.128 and 4.226 GeV with the BEPCII collider and collected with the BESIII detector.  Charge-conjugate modes are implied throughout this Letter.

Details of the BESIII detector are described in Ref.~\cite{Ablikim2010345}. About 83\% of the data analyzed in this Letter benefits from an upgrade of the end cap time-of-flight system
with multi-gap resistive plate chambers yielding a time resolution of 60\,ps~\cite{Lixin,Gyingxiao}. Monte Carlo (MC) events are generated with a {\sc{geant4}}-based~\cite{Agostinelli:2002hh} software package, which includes the geometric description of the BESIII detector~\cite{Huang:2022wuo} and the detector response. An inclusive MC sample corresponding to 40 times the integrated luminosity of the data is produced at $E_{\rm CM}$ between 4.128 and 4.226~GeV. It includes open charm processes, initial state radiation (ISR) production of charmonium [$\psi(3770)$, $\psi(3686)$ and $J/\psi$], $q\bar q\,(q=u,\,d,\,s$) continuum processes, as well as Bhabha scattering, $\mu^+\mu^-$, $\tau^+\tau^-$ and $\gamma\gamma$ events.  The open charm processes are generated using {\sc{conexc}}~\cite{Ping:2013jka}. The effects of ISR and final state radiation are included.  Signal MC samples of the SL decay $D^+_s \to f_{0}(980) \mu^+\nu_\mu$ with $f_{0}(980) \to \pi^+ \pi^-$ are simulated with the simple pole parameterization model~\cite{FF-simple}, using parameters determined in this work.  The amplitude for the $f_{0}(980)$ is modeled using the Flatté formula with parameters fixed to the BESII measurement~\cite{bes2f980}. All particle decays are modeled with {\sc{evtgen}}~\cite{ref:evtgen} using BFs either taken from the Particle Data Group~\cite{pdg}, when available, or otherwise estimated with {\sc{lundcharm}} \cite{ref:lundcharm}. Final state radiation from charged final state particles is incorporated using the {\sc photos} package~\cite{PHOTOS}.

In $e^+e^-$ collisions at $E_{\rm CM}$ between 4.128 and 4.226~GeV, the $D_s^+$ mesons are produced predominantly via $e^+e^-\to D^\pm_sD_s^{*\mp}$.
Candidates in which one $D_s^-$ is fully reconstructed in one of the fifteen hadronic decay modes,
$D^-_s\to K^+K^-\pi^-$, $K^-\pi^+\pi^-$, $K^+K^-\pi^-\pi^0$, $K^0_SK^-$,
$K^0_SK^-\pi^0$,
$K^0_SK^0_S\pi^-$,
$K^0_SK^+\pi^-\pi^-$,
$K^0_SK^-\pi^+\pi^-$,
$\pi^+\pi^-\pi^-$,
$\eta_{\gamma\gamma}\pi^-$, $\eta_{\pi^0\pi^+\pi^-}\pi^-$,
$\eta^\prime_{\eta_{\gamma\gamma}\pi^+\pi^-}\pi^-$, $\eta^\prime_{\gamma\rho^0}\pi^-$,
$\eta_{\gamma\gamma}\rho^-$, and
$\eta_{\pi^+\pi^-\pi^0}\rho^-$,
 are referred to as single-tag (ST) candidates. Those in which the ST $D^-_s$, the transition $\gamma(\pi^0)$ of the $D_s^{*\mp}$ decay, and the signal $D^+_s$ decay of interest are simultaneously reconstructed are referred to as double-tag~(DT) events. Based on these, we determine the BF of the signal decay by
\begin{equation}
{\mathcal B}_{\rm sig} = N_{\rm DT}/(N^{\rm tot}_{\rm ST} \, \epsilon_{\gamma(\pi^0)\rm sig}),\label{eq:branch}
\end{equation}
where $N^{\rm tot}_{\rm ST}=\sum_k N^k_{\rm ST}$ and $N_{\rm DT}$ are the ST and DT yields in data, and $\epsilon_{\gamma(\pi^0)\rm sig}$ is the effective signal efficiency of selecting $\gamma(\pi^0)f_{0}(980)\mu^+\nu_\mu$ in the presence of ST $D^-_s$. The $\epsilon_{\gamma(\pi^0)\rm sig}$ is calculated by $\sum_k\frac{N^k_{\rm ST}}{N^{\rm tot}_{\rm ST}}\frac{\epsilon^k_{\rm DT}}{\epsilon^k_{\rm ST}}$, where $\epsilon^k_{\rm ST}$ and $\epsilon^k_{\rm DT}$ are the ST and DT efficiencies for the $k^{th}$ tag mode, respectively. 

For each tag mode, the ST yield is extracted from a fit to the corresponding invariant-mass spectrum of the ST candidates. The selection criteria for all ST candidates except $D^-_s\to \eta_{\pi^+\pi^-\pi^0}\rho^-$ are the same as Ref.~\cite{Ablikim:2024prd092003}, where a detailed description can be found. For the tag modes $D^-_s\to \eta_{\pi^0\pi^+\pi^-}\rho^-$, the $\pi^0\pi^+\pi^-$ decay mode is used to form $\eta$ candidates and the invariant mass, $M_{\pi^0\pi^+\pi^-}$, is required to be within the interval $(0.530,\,0.570)~\mathrm{GeV}/c^2$. The $\rho^-$ candidates are reconstructed from $\pi^-\pi^0$ combinations with invariant mass within the interval $(0.570,\,0.970)~\mathrm{GeV}/c^2$. Summing over all tag modes, we obtain a total ST yield $N^{\rm tot}_{\rm ST}=(839.1\pm3.7_{\rm stat}\pm4.2_{\rm syst})\times 10^3$.

In the presence of ST $D_s^-$, we select candidates for the transition $\gamma\,(\pi^0)$ from $D_s^{*\mp}$ decay and signal $D_s^+\to f_{0}(980)\mu^+\nu_\mu$ among the unused particles recoiling against the ST $D_s^-$. The $f_{0}(980)$ is reconstructed via the $f_{0}(980)\to\pi^+\pi^-$ decay. Particle identification (PID) for pions combines the specific ionization information from the multilayer drift chamber and
the flight time in the time-of-flight system, and PID for muons further combines the energy deposited in the electromagnetic calorimeter~(EMC). Likelihoods under various particle hypotheses (${\mathcal L}_i$, $i=e$, $\pi$, $\mu$, and $K$) are calculated. Charged tracks satisfying ${\mathcal L}_\pi>0.001$, ${\mathcal L}_\pi>{\mathcal L}_K$ are assigned as pion candidates, and those satisfying ${\mathcal L}_\mu>0.001$, ${\mathcal L}_\mu>{\mathcal L}_e$, ${\mathcal L}_\mu>{\mathcal L}_K$ and $E_{\rm EMC} \in (0.10, 0.28)$~GeV are assigned as muon candidates, where $E_{\rm EMC}$ is the energy deposited in the EMC by muon candidates.
The selection criteria for the transition $\gamma\,(\pi^0)$ of the $D_s^{*\mp}$ decay are the same as those in Refs.~\cite{Ablikim:2024prd092003,Ablikim:2024prl091802}.  PID is unable to fully distinguish $\mu^+$ and $\pi^+$ tracks leading to some misidentifications.  To minimize the efficiency loss, both of the $+-$ track pairings are considered as $f_0(980)$ decay products.  That is, both ($\mu^+\pi^-$)$\pi^+$ and ($\pi^+\pi^-$)$\mu^+$ candidate combinations are retained, where the candidates in (...) indicate the $f_0(980)$ decay products. To  suppress backgrounds from hadronic $D_s$ decays, the maximum energy of any photon not used in the SL selection, $E^{\rm max}_{\rm extra~ \gamma}$, is required to be less than 0.20~GeV, and the number of extra unused charged tracks, $N^{\rm char}_{\rm extra}$, and extra $\pi^0$ candidates from two unused photons, $N^{\pi^{0}}_{\rm extra}$, must both be zero.   Here, the selection criteria for photon, charged tracks, and $\pi^0$ are the same as Refs.~\cite{Ablikim:2024prd092003,Ablikim:2024prl091802}.
 
 The energy, $E_{\rm miss}$, and momentum, $\vec{p}_{\rm miss}$, of the missing neutrino are estimated using energy and momentum conservation. They are calculated in the $e^+e^-$ rest frame as $E_{\rm miss}\equiv E_{\rm CM}-\sum_{j}E_{j}$ and $\vec{p}_{\rm miss} \equiv -\sum_{j}\vec{p}_{j}$. The index $j$ sums over the ST $D_s^-$, transition $\gamma\,(\pi^0)$, $\pi^+\pi^-$ pair, and $\mu^+$, where $E_j$ and $\vec{p}_j$ are the energy and momentum of the $j^{th}$ particle, respectively. The missing neutrino is characterized by the missing mass squared, $M^{2}_{\rm miss} \equiv E^{2}_{\rm miss}/c^4-{|\vec{p}_{\rm miss}|}^2/c^2$. To improve the $M^{2}_{\rm miss}$ resolution, the candidate particles, together with the missing neutrino, are subjected to a 3-constraint kinematic fit requiring energy and momentum conservation, constraining the invariant mass of each  $D_s^\pm$ meson to the known $D_s^\pm$ mass~\cite{pdg}, and constraining the invariant mass of the $D_s^-\gamma(\pi^0)$ or $D_s^+\gamma(\pi^0)$ combination to the known $D_s^{*\pm}$ mass~\cite{pdg}. The combination with the lowest $\chi^2_{3C}$ is kept. To further suppress non-$D_s^\pm D_s^{*\mp}$ backgrounds, we require $-0.25~\text{GeV}^2/c^4 < M^2_{\text{miss}} < 0.15~\text{GeV}^2/c^4$ and  $\chi^2_{3C}< 20$.

Background from $D_s^+\to\pi^+\pi^- \pi^+$ is suppressed by requiring the $f_0(980)\mu^+$ invariant mass to satisfy $M_{f_0(980)\mu^+} <$ 1.72~GeV/$c^2$.  To suppress the background  from $D_s^+\to \eta^{'}_{\gamma \rho}\pi^+$, we require that the $f_0(980)\nu_\mu$ invariant mass satisfy $M_{f_0(980)\nu_\mu} >$ 1.22~GeV/$c^2$. To reject the backgrounds from $D_s^+\to\pi^+\pi^-e^+\nu_e$, we require $E_{\pi^+}/(p_{\pi^+}c)  < 0.85$, where $E_{\pi^+}$ is the energy deposited by the $\pi^{+}$ meson, and $p_{\pi^+}$ is the momentum of the $\pi^{+}$ meson.
Additionally, to further reject backgrounds from hadronic  $D_s^+$ decays,  the opening angle between the missing momentum and the most energetic unused shower, $\theta_{\vec{p}_{\rm miss}, \gamma}$, is required to satisfy $\cos\theta_{\vec{p}_{\rm miss}, \gamma} <$ 0.70.

\begin{figure}[htp]
\begin{center}
\includegraphics[width=\linewidth]{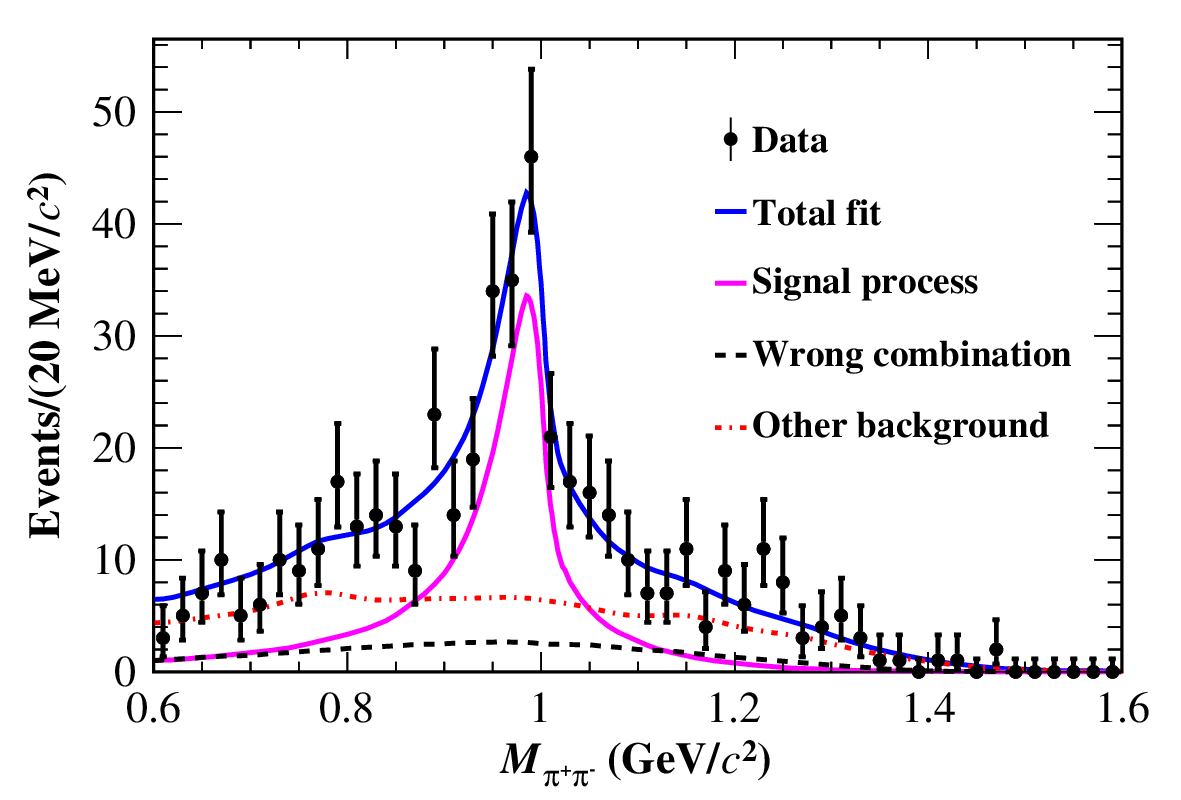}
\caption{ Fit to the ${M_{\rm \pi^{+} \pi^{-}}}$ distribution of the candidate events for $D_{s}^{+} \to f_0(980)\mu^+\nu_{\mu}$.
Points with error bars represent data. he solid blue curve denotes the total fit. The solid violet curve shows the signal process.
The long-dashed black and dash-dotted red curves are the background contributions from the wrong combinations from signal events and other combinatorial backgrounds, respectively. 
}
\label{fit:hadMass-data}
\end{center}
\end{figure}

To reconstruct the $f_{0}(980)$, we require the $\pi^+\pi^-$ invariant mass ($M_{\pi^+\pi^-}$) to be within the interval (0.6, 1.6) ${\rm GeV}/c^{2}$, and the resulting $M_{\pi^+\pi^-}$ distribution of the accepted candidates is shown in Fig.~\ref{fit:hadMass-data}. To obtain the signal yield, an unbinned maximum likelihood fit is performed on the combined spectra of $M_{\pi^+\pi^-}$  for the two possible $+-$ track pairings. In the fit, the signal is modeled by the MC-simulated shape for events with the matching angle between the reconstructed and generated four-momentum of the $\pi^+$, $\pi^-$, and $\mu^+$ in the signal decay all less than $15^\circ$.  This shape is then convolved with a Gaussian smearing function, with free parameters representing the data-MC resolution difference. The wrong combinations from signal MC are treated as a background,  using the MC-simulated shape for candidates with a matching angle greater than $15^\circ$ for any track. The wrong-combination yield is fixed at $0.343 N_{\rm DT}$; the pre-factor is the ratio of signal MC events with any track matching angle greater than $15^\circ$ to those with all matching angles less than $15^\circ$. The total yield of all other backgrounds from inclusive MC  is floated. Both of these background shapes are smeared with the same Gaussian function as used for the signal. 
From the fit, shown in Fig.~\ref{fit:hadMass-data}, we obtain $196\pm22$ signal events. The averaged detection efficiency $\epsilon_{\gamma(\pi^0)\rm sig}$ is estimated to be  $(14.72\pm0.08)\%$. Using  Eq.~(\ref{eq:branch}), the corresponding BF is determined to be $\mathcal{B}(D^+_s\to f_0(980)\mu^+\nu_\mu, f_0(980)\to\pi^+\pi^-) = (1.59\pm0.18_{\rm stat}\pm0.11_{\rm syst})\times 10^{-3}$, where the uncertainties are statistical and systematic, respectively. Using the branching fraction ${\mathcal B}[f_0(980) \to \pi^+ \pi^-] = (46 \pm 6)\%$, assuming $\pi\pi$ and $KK$ decays dominate~\cite{f0980pipi}, along with the relation ${\mathcal B}[D_s^+ \to f_0(980) \mu^+\nu_\mu] = (4.42 \pm 2.0) \times 10^{-3} \cos^2 \phi$~\cite{QCDSR2}, where the mixing angle $\phi$ contributes to the $(q\bar{q})$ mixed picture of $f_0(980)$ represented as $\sin \phi \left(1/\sqrt{2}\right)(u\bar{u} + d\bar{d}) + \cos \phi s\bar{s}$~\cite{QCDSR2,LF-RQM}, we obtain the angle $\phi = (28.0 \pm 26.6)^\circ$, indicating that the $s\bar{s}$ component is likely dominant. 

The systematic uncertainties in the BF measurement are discussed next. The uncertainty from the ST $D_s^-$ yield is determined to be 0.5\%~\cite{Ablikim:2024prl091802,Ablikim:2024prd092003}.  The efficiencies of $\mu^+$ and $\pi^{\pm}$ tracking and PID are verified using $e^+e^-\to \gamma\mu^+\mu^-$ events and $e^+e^-\to K^+K^-\pi^+\pi^-$ hadronic events, respectively. We assign the uncertainties of  $\mu^+$  tracking (PID) and $\pi^+\pi^-$ pair tracking (PID) to be 0.3\% (0.4\%) and 0.6\% (0.4\%), respectively~\cite{Ablikim:2024prl091802,Ablikim:2024prd092003}. The uncertainty in the transition $\gamma(\pi^0)$ reconstruction is estimated to be 1.0\% by analyzing the control sample $J/\psi\to \pi^0\pi^+\pi^-$~\cite{Ablikim:2011kv}. The uncertainty due to different multiplicities of tag environments is 0.3\% by studying the differences between data-MC efficiencies~\cite{Ablikim:2018jun}. The uncertainties associated with the requirements on $\cos\theta_{\vec{p}_{\rm miss}, \gamma}$, $M^2_{\text{miss}}$, and $\chi^2_{3C}$ are estimated to be 1.5\%, 3.8\%, and 2.7\%, respectively, using the DT sample $D_s^+\to K^+K^-e^+\nu_e$. The total uncertainty associated with the $E^{\rm max}_{\rm extra\gamma}$, $N^{\pi^{0}}_{\rm extra}$, and $N^{\rm char}_{\rm extra}$  requirements is estimated to be 0.4\%  by using the control sample of $D_s^+\to K^+K^-\pi^+$. The uncertainty associated with the $M_{f_0(980)\mu^+} $ and $M_{f_0(980)\nu_\mu}$ requirements is estimated to be 2.7\% by analyzing the DT events of $D_s^+\to\eta^{\prime}_{\pi^+\pi^-\gamma}e^+\nu_e$. The uncertainty associated with the fit to the $M_{\pi^+\pi^-}$ distribution is estimated to be 1.3\% arising from the signal shape, background shape, and ratio of wrong signal combinations as follows. 
The systematic uncertainty from the signal shape is determined by repeating the fits, treating both the $\mu^+\pi^-$ and $\pi^+\pi^-$ candidate combinations as the signal events, and considering the BF difference as the systematic uncertainty for this item. The systematic uncertainty due to the background shape is studied by varying the relative fractions of the major backgrounds from $e^+e^- \to q\bar{q}$ and the non-$D_s^{*+}D^{-}_s$ open charm processes by $\pm30\%$, based on the errors  of input cross sections in the inclusive MC sample~\cite{bes3f0980}. The systematic uncertainty from the ratio of wrong signal combinations is studied by changing the matching angle from $15^\circ$ to $5^\circ$, $10^\circ$, $20^\circ$, and $25^\circ$ to evaluate the impact on the BF; the largest BF difference is taken as the systematic uncertainty.  The uncertainty due to the selection of the transition $\gamma(\pi^0)$ from $D_s^{*+}$ is estimated to be 1.0\% by using the control samples of $D_s^+\to K^+K^-\pi^+$ and $D_s^+\to \eta\pi^0\pi^+$~\cite{Ablikim:2024prl091802}. The uncertainty due to limited MC statistics is 0.5\%, dominated by that of the DT efficiency, and  is taken as a systematic uncertainty. The uncertainty related to the signal MC model is estimated to be 2.8\% by replacing the $f_0(980)$ line shape from BESII~\cite{bes2f980} with the one from LHCb~\cite{lhcb2012} in generating the signal MC samples. The uncertainty due to the quoted BFs of $D_s^{*+}$ decays is 0.3\%~\cite{pdg}. This uncertainty is determined by examining the variation of the signal efficiency when varying the quoted BFs of $D_s^{*+}\to \gamma D_s^+$ and $D_s^{*+}\to \pi^0 D_s^+$ within $\pm 1\sigma$. The systematic uncertainty contributions are summed in quadrature  to obtain the total systematic uncertainty of 6.7\%.

The dynamics of $D_{s}^{+}\to f_{0}(980)\mu^+\nu_{\mu}$ decay is studied by dividing the SL candidate events into four intervals of $q^2$ (four-momentum transfer square of $\mu^{+}\nu_{\mu}$). Using the measured and expected partial decay rates of the $i^{th}$ $q^{2}$ interval, $\Delta\Gamma^{i}_{\rm mea}$ and $\Delta\Gamma^{i}_{\rm exp}$, the FF $f^{f_{0}(980)}_{+}(q^2)$ is determined by constructing and minimizing a $\chi^2$ as 
\begin{equation}
\chi^{2}=\sum_{ij} (\Delta\Gamma^{i}_{\rm mea} - \Delta\Gamma^{i}_{\rm exp}) \, C_{ij}^{-1} \, (\Delta\Gamma^{j}_{\rm mea} - \Delta\Gamma^{j}_{\rm exp}),
\label{eq:chi}
\end{equation}
where $C_{ij}$ is the covariance matrix that accounts for correlations of $\Delta\Gamma^{i}_{\rm mea}$ among $q^2$ intervals.

The $\Delta\Gamma^{i}_{\rm exp}$ is calculated by integrating the following double differential decay rate~\cite{FF-rate-sq0,FF-rate-sq1,FF-rate-sq2}

\begin{widetext}
\begin{equation}
\begin{split}
\frac{d^2\Gamma}{ds dq^2} \,=\, \frac{G^2_F |V_{cs}|^2}{192\pi^4 m_{D_s^+}^3} \, \lambda^{1/2}\left( m_{D_s^+}^2,s, q^2 \right) \bigg(1-\frac{m_\mu^2}{q^2}\bigg)^2
\Bigg[ & \bigg( 1+\frac{m_\mu^2}{2q^2} \bigg) \lambda\left(m_{D_s^+}^2, s , q^2\right) \left|f^{f_{0}(980)}_+(q^2)\right|^2 \\
& + \frac{3m_\mu^2}{2q^2} \left(m_{D_s^+}^2- s \right)^2 \left|f^{f_{0}(980)}_0(q^2)\right|^2 \Bigg] P(s),
\end{split}
\end{equation}
\end{widetext}
where $s$ is the square of $M_{\pi^+\pi^-}$, $G_F$ is the Fermi constant~\cite{pdg}, $|V_{cs}|$ is the modulus of the Cabibbo Kobayashi-Maskawa matrix element, $m_{D^+_s}$ is the known $D^+_s$ mass~\cite{pdg},
$\lambda(x,y,z)=x^2+y^2+z^2-2xy-2xz-2yz$, and $P(s)$ is based on the relativistic Flatt\'e formula~\cite{bes2f980} with the open $K^+K^-$:
\begin{equation}
P(s)=\frac{g_1\rho_{\pi\pi}}{|m^2_0-s-i(g_1\rho_{\pi\pi} + g_2\rho_{K\bar{K}})|^2}.
\label{eq:Ps}
\end{equation}
Here $m_0$ denotes the $f_0(980)$ mass, the constants $g_1$ and $g_2$ are the $f_0(980)$ couplings to $\pi^+\pi^-$ and $K^+K^-$ final states, respectively, and $\rho_{\pi\pi}$ and $\rho_{K\bar{K}}$ are individual phase space factors. The function $P(s)$ describes the line shape of the $f_0(980)$ resonance in the $M_{\pi^+\pi^-}$ spectrum. The hadronic transition FF, $f^{f_{0}(980)}_{+}(q^2)$, is modeled with the simple pole parameterization~\cite{FF-simple}, in which the product $f^{f_{0}(980)}_{+}(0)|V_{cs}|$ and the pole mass $M_{\rm pole}$ are extracted from the fit. The same formula is applied for $f^{f_{0}(980)}_{0}(q^2)$, but the contribution from this term is much smaller.

The measured partial decay rate $\Delta\Gamma_{\rm mea}^{i}$ is determined by $\Delta\Gamma_{\rm mea}^{i} \equiv \int_{i}\int_{s} \frac{d^{2}\Gamma}{ds \, dq^{2}} \, ds \, dq^{2}=N^i_{\rm pro}/(\tau N^{\rm tot}_{\rm ST})$,
where $\tau$ is the $D_{s}^{+}$ meson lifetime~\cite{pdg,lifetime} and $N_{\rm pro}^{i}$ is the SL signal yield produced in the $i^{th}$ $q^2$ interval, obtained as $N_{\rm pro}^{i}=\sum_{j=1}^{4} \epsilon^{-1}_{ij} N_{\rm obs}^{j}$. Here $N_{\rm obs}^{j}$ is the observed SL decay yield obtained from the similar fit to the corresponding $M_{\pi^{+}\pi^{-}}$ distribution as described previously for the BF measurement, and $\epsilon_{ij}$ is the efficiency matrix determined from the signal MC samples via $\epsilon_{ij} = \sum_{k}[(1/N_{\rm tag}^{\rm tot}) \, (N_{\rm rec}^{ij}/N_{\rm gen}^{j})_{k} \, (N_{\rm tag}^{k}/\epsilon_{\rm tag}^{k})]$, where $N_{\rm rec}^{ij}$ is the SL decay yield reconstructed in the $i^{th}$ $q^2$ interval and generated in the $j^{th}$ $q^2$ interval, $N_{\rm gen}^{j}$ is the total signal yield generated in the $j^{th}$ $q^2$ interval, and $k$ sums over all tag modes. The details of the divisions, $N_{\rm pro}^{i}$, $N_{\rm obs}^{i}$, and $\Delta\Gamma_{\rm mea}^{i}$ of various $q^2$ intervals are given  in Table 1 of Ref.~\cite{Supplement}.

The statistical and systematic covariance matrices are constructed as $C_{ij}^{\rm stat}=[1/(\tau N_{\rm tag}^{\rm tot})^{2}]\sum_{m}\epsilon_{im}^{-1}\epsilon_{jm}^{-1}[\sigma(N^{m}_{\rm obs})]^{2}$ and $C_{ij}^{\rm syst}=\delta(\Delta\Gamma_{\rm mea}^{i})\delta(\Delta\Gamma_{\rm mea}^{j})$, respectively, where $\sigma(N^{m}_{\rm obs})$ and $\delta(\Delta\Gamma_{\rm mea}^{i})$ are the statistical and systematic uncertainties in the $m^{th}$ and $i^{th}$ $q^2$ intervals~\cite{Ablikim:2024prl091802}. The $C_{ij}^{\rm syst}$ is obtained by summing all the covariance matrices for all systematic uncertainties, where the systematic uncertainty on $\tau$, 0.8\%~\cite{pdg,lifetime}, is included along with those for the BF measurement. The correlation matrices of the obtained $C_{ij}^{\text{stat}}$ and $C_{ij}^{\text{syst}}$ are shown in Tables 2, 3 in Ref.~\cite{Supplement}. 

The systematic uncertainty related to $\Delta\Gamma_{\rm mea}^{i}$ is estimated to be 4.1\% by following Ref.~\cite{bes3f0980}. In addition, the uncertainties of the input parameters $m_0$, $g_1$, and $g_2$~\cite{bes2f980} related to $\Delta\Gamma^{i}_{\rm exp}$ are also considered by varying them by $\pm 1 \sigma$ from their central values. The largest deviations of the FF, namely 1.2\%, 0.7\%, and 2.4\%, are taken as systematic uncertainties. The quadrature sum of the above uncertainties is 5.0\%, which is taken as the total systematic uncertainty. 

\begin{figure*}[htp]
\begin{center}
\includegraphics[width=0.80\textwidth]{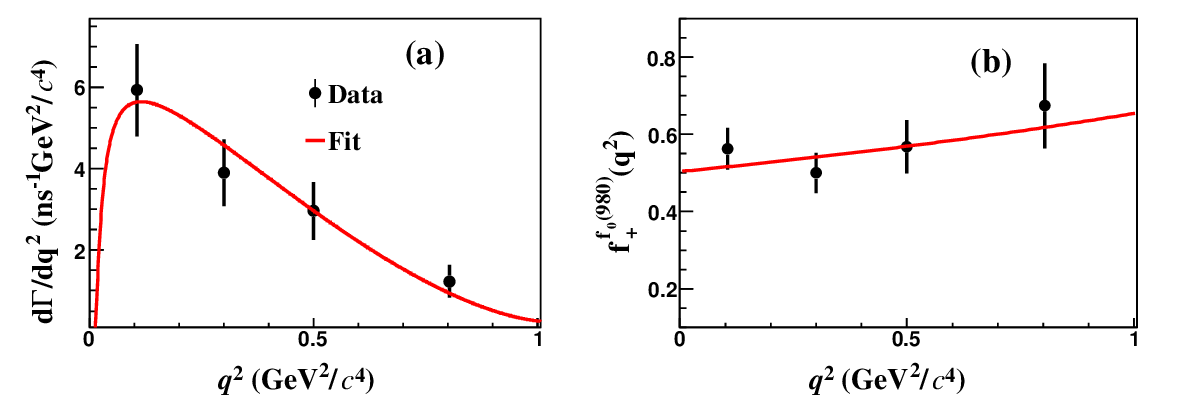}
\caption{Shown are (a) a fit the differential decay rate as a function of $q^2$ and (b) the corresponding FF $f^{f_{0}(980)}_{+}(q^2)$. 
The points with error bars are data, and the red line is the fit.}
\label{fit:ff-width}
\end{center}
\end{figure*}

A fit is performed for the differential decay rate of the channel $D_{s}^{+}\to f_{0}(980) \mu^+\nu_\mu$ and the FF projection are shown in Fig.~\ref{fit:ff-width}. The product of the FF and $|V_{cs}|$ is determined to be $f^{f_{0}(980)}_{+}(0)|V_{cs}|=0.490\pm0.059_{\rm stat}\pm0.025_{\rm syst}$. The fit result is shown in Fig.~\ref{fit:ff-width}(a), while Fig.~\ref{fit:ff-width}(b) shows the same fit in projection to the FF $f^{f_{0}}_{+,0}(q^2)$.

In summary, using $e^+e^-$ collision data corresponding to an integrated luminosity of 7.33${~\rm fb^{-1}}$ collected at $E_{\rm CM}\in [4.128,4.226]$~GeV by the BESIII detector, we report the observation of the decay $D^+_s\to f_0(980) \mu^+\nu_\mu$ and the study of its dynamics for the first time. The BF of $D^+_s\to f_0(980)\mu^{+}\nu_{\mu},~f_0(980)\to\pi^+\pi^-$ decay is measured to be $(1.59\pm0.18_{\rm stat}\pm0.11_{\rm syst})\times 10^{-3}$, which implies that the $s\bar{s}$ component dominates for the $f_{0}(980)$ in the $q\bar{q}$ mixture picture. Combining with the BESIII measurement of ${\mathcal B}(D^+_s\to f_{0}(980) e^+\nu_e)$~\cite{bes3f0980}, the ratio is determined to be $\frac{{\mathcal B}(D^+_s\to f_{0}(980) \mu^+\nu_\mu)}{{\mathcal B}(D^+_s\to f_{0}(980)e^+\nu_e)} = 0.92\pm0.13_{\rm stat}\pm0.08_{\rm syst}$, where the systematic uncertainties from $\pi^+\pi^-$ pair tracking and PID have been cancelled. No evidence for LFU violation is observed with the current precision.

\begin{figure}
\centering
  \includegraphics[width=0.475\textwidth]{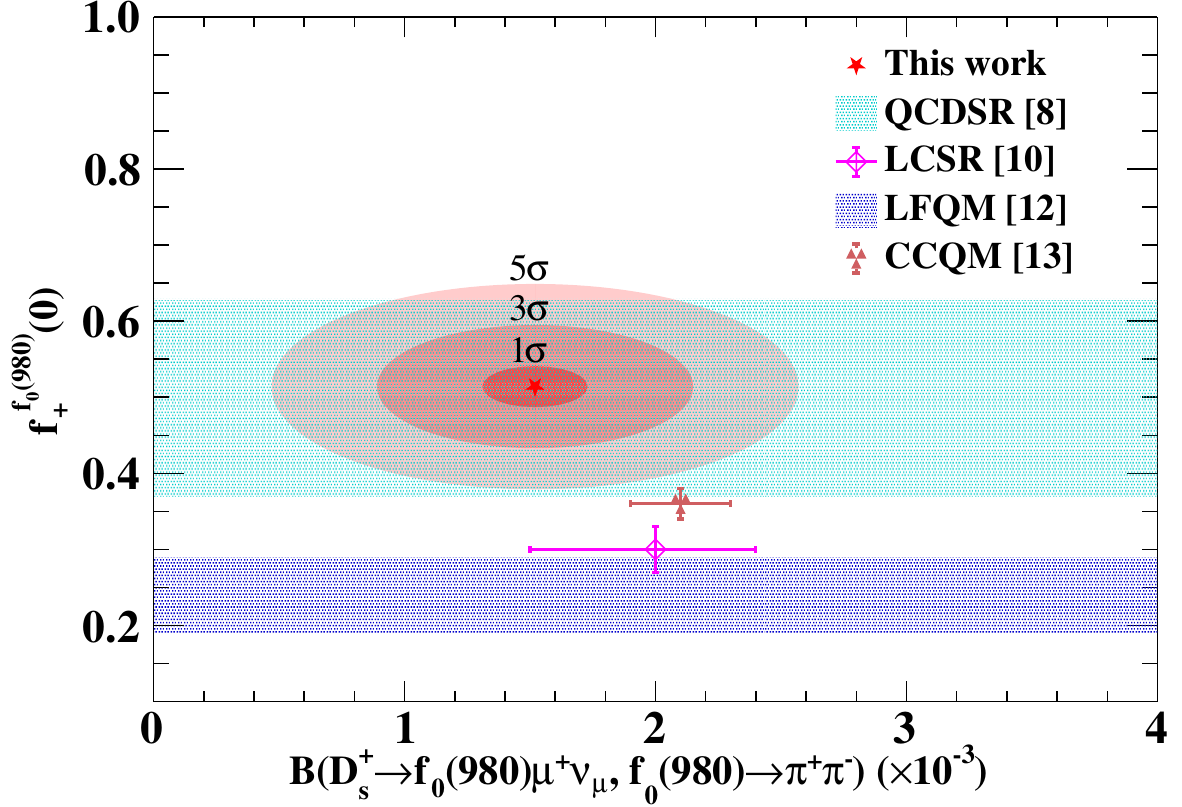}
\caption{ Comparison of the measured BF and FF $f^{f_{0}(980)}_{+}(0)$ with several theoretical predictions.}
  \label{fig:BF_FF_compare}
\end{figure}

By analyzing the $D^+_s\to  f_0(980) \mu^+\nu_\mu$ decay dynamics, we obtain the product $f^{f_{0}(980)}_{+}(0)|V_{cs}|=0.490\pm0.059_{\rm stat}\pm0.025_{\rm syst}$. Averaging with our previously reported result for the $D_{s}^{+} \to f_{0}(980)e^+\nu_e$ decay, we determine $f^{f_{0}(980)}_{+}(0)|V_{cs}|=0.500\pm0.016_{\rm stat}\pm0.020_{\rm syst}$. Using $|V_{cs}|=0.97349\pm0.00016$ from the CKMfitter group~\cite{pdg} as input, we obtain the most precise FF $f^{f_{0}(980)}_{+}(0)=0.514\pm0.017_{\rm stat}\pm0.021_{\rm syst}$ for $D_{s} \to f_{0}(980)$ transition with the simple pole parameterization. 
Figure~\ref{fig:BF_FF_compare} compares our measured BF and FF $f^{f_{0}(980)}_{+}(0)$ with various theoretical predictions. Our measured BF agrees within uncertainties with predictions from the light-cone QCD sum rules (LCSR) and covariant confined quark model (CCQM)~\cite{LCSR1,LCSR2,Soni2020}. For the FF $f^{f_{0}(980)}_{+}(0)$, our results are consistent with calculations from the covariant light-front dynamics (CLFD), dispersion relation (DR), and QCD sum rule (QCDSR) frameworks~\cite{qqbar2,QCDSR1,QCDSR2}. By contrast, the measured FF $f^{f_{0}(980)}_{+}(0)$ value exhibits a significant discrepancy of up to 5$\sigma$ with the LCSR, light-front quark model (LFQM), and CCQM~\cite{LCSR1,LF-RQM,Soni2020} predictions, clearly disfavoring these theoretical models. Notably, all theoretical predictions considered here have incorporated the finite width of $f_{0}(980)$ in their FF $f^{f_{0}(980)}_{+}(0)$ computations. Our measurements offer essential input for advancing the investigation of the intrinsic properties of the light scalar state $f_0(980)$, as well as the decay dynamics of charm mesons in the non-perturbative QCD regime.

\begin{acknowledgments}
The authors thank Prof. Haibing Fu for helpful discussions. The BESIII Collaboration thanks the staff of BEPCII (https://cstr.cn/31109.02.BEPC) and the IHEP computing center for their strong support. This work is supported in part by National Key R\&D Program of China under Contracts Nos. 2023YFA1606000, 2023YFA1606704, 2025YFA1613900; National Natural Science Foundation of China (NSFC) under Contracts Nos. 12375092, 11635010, 11935015, 11935016, 11935018, 12025502, 12035009, 12035013, 12061131003, 12192260, 12192261, 12192262, 12192263, 12192264, 12192265, 12221005, 12225509, 12235017, 12342502, 12361141819, 12535005; the Chinese Academy of Sciences (CAS) Large-Scale Scientific Facility Program; the Strategic Priority Research Program of Chinese Academy of Sciences under Contract No. XDA0480600; CAS under Contract No. YSBR-101; 100 Talents Program of CAS; The Institute of Nuclear and Particle Physics (INPAC) and Shanghai Key Laboratory for Particle Physics and Cosmology; ERC under Contract No. 758462; German Research Foundation DFG under Contract No. FOR5327; Istituto Nazionale di Fisica Nucleare, Italy; Knut and Alice Wallenberg Foundation under Contracts Nos. 2021.0174, 2021.0299, 2023.0315; Ministry of Development of Turkey under Contract No. DPT2006K-120470; National Research Foundation of Korea under Contract No. NRF-2022R1A2C1092335; National Science and Technology fund of Mongolia; Polish National Science Centre under Contract No. 2024/53/B/ST2/00975; STFC (United Kingdom); Swedish Research Council under Contract No. 2019.04595; U. S. Department of Energy under Contract No. DE-FG02-05ER41374.

\end{acknowledgments}

\clearpage
\appendix
\twocolumngrid
\setcounter{table}{0}
\setcounter{figure}{0}
\section*{Supplemental Material}\label{Supplement}

Tables~\ref{tab:frac-width} presents the number of reconstructed events $N^{i}_{\rm sig}$ in data obtained from the $M_{\pi^-\pi^+}$ fits, the number of produced events $N^i_{\rm pro}$, and the measured partial decay rate $\Delta\Gamma^i_{\rm mea}$  in different $q^2$ intervals, for $D^+_s\to f_0(980)\mu^+\nu_\mu$ with $f_0(980)\to\pi^+\pi^-$.

Table~\ref{tab:rho_stat} summarizes the statistical correlation matrix for the measured partial decay rates in different  $q^2$ intervals, for $D^+_s\to f_0(980)\mu^+\nu_\mu$ with $f_0(980)\to\pi^+\pi^-$.

Table~\ref{tab:rho_sys} summarizes the systematic correlation matrix for the measured partial decay rates in different  $q^2$ intervals, for $D^+_s\to f_0(980)\mu^+\nu_\mu$ with $ f_0(980)\to\pi^+\pi^-$.

\begin{table}[htp]
\centering
\caption[Fraction Width]{The partial decay rates in different $q^2$ intervals of data, with statistical uncertainties only.}
\vspace{0.25cm}
\begin{tabular}{ccccc}
\hline
\hline
$i$ & 1 & 2 & 3 & 4 \\
$q^2~({\rm GeV}^2/c^4)$ & ($m^{2}_{\mu}$, 0.20) & (0.20, 0.40) & (0.40, 0.60) & (0.60, $q^2_{\rm max}$) \\
\hline
$N^{i}_{\rm sig}$       & $69\pm12$ & $57\pm11$ & $47\pm11$ & $34\pm10$ \\
$N^{i}_{\rm pro}$       & $482\pm86$ & $329\pm69$ & $250\pm60$ & $197\pm63$ \\
$\Delta\Gamma^i_{\rm mea}~({\rm ns}^{-1})$ & $1.14\pm0.20$ & $0.78\pm0.16$ & $0.59\pm0.14$ & $0.47\pm0.15$ \\
\hline
\hline
\end{tabular}
\label{tab:frac-width}
\end{table}

\begin{table}[htp]
\centering
\caption{Statistical correlation matrix for the measured partial decay rates of $D^+_s\to f_0(980)\mu^+\nu_\mu,f_0(980)\to\pi^+\pi^-$ in different $q^2$ intervals.}
\vspace{0.25cm}
\begin{tabular}{cccccc}
\hline
\hline
$\rho_{ij}^{\text{stat}}$               & 1            & 2           & 3            & 4             \\
\hline
1                        &1.000  &-0.047 &0.002  &-0.000             \\
2                        & &1.000  &-0.060 &0.002             \\
3                        &     & &1.000  &-0.059               \\
4                        &    &  &  &1.000            \\
\hline\hline
\end{tabular}
\label{tab:rho_stat}
\end{table}

\begin{table}[htp]
\centering
\caption{Systematic correlation matrix for the measured partial decay rates of $D^+_s\to f_0(980)\mu^+\nu_\mu,f_0(980)\to\pi^+\pi^-$ in different $q^2$ intervals.}
\vspace{0.25cm}
\begin{tabular}{ccccccccc}
\hline
\hline
$\rho_{ij}^{\text{syst}}$                 & 1            & 2           & 3            & 4           \\
\hline
1       &1.000  &0.877  &0.914  &0.364  \\
2       &      &1.000  &0.884  &0.458  \\
3       &       &  &1.000  &0.414  \\
4       &      &   &   &1.000  \\
\hline\hline
\end{tabular}
\label{tab:rho_sys}
\end{table}

\end{document}